\documentclass[twoside]{article}
\usepackage{fleqn,espcrc2}

\newcommand{\AmS}{{\protect\the\textfont2
  A\kern-.1667em\lower.5ex\hbox{M}\kern-.125emS}}

\newcommand{\be}{\begin{equation}}\newcommand{\ee}{\end{equation}}
\newcommand{\bea}{\begin{eqnarray}}\newcommand{\eea}{\end{eqnarray}}
\newcommand{\nn}{\nonumber}\newcommand{\p}[1]{(\ref{#1})}
\newcommand{\lb}[1]{\label{#1}}
\newcommand\s{\scriptscriptstyle}

\newcommand\q{\quad}

\newcommand{\vp}{\varphi}
\newcommand{\bvp}{{\bar\varphi}}
\newcommand{\bnu}{{\bar\nu}}

\newcommand\cA{{\cal A}}

\newcommand\cL{{\cal L}}

\newcommand\cS{{\cal S}}

\newcommand{\da}{{\dot{\alpha}}}
\newcommand{\db}{{\dot{\beta}}}

\newcommand\ab{{\alpha\beta}}

\newcommand\adb{{\alpha\db}}
\newcommand\ada{{\alpha\da}}

\newcommand\padb{\partial_\adb}

\newcommand\B{{\s B}}

\newcommand\D{{\s D}}
\newcommand\I{{\s I}}

\title{NEW REPRESENTATION FOR LAGRANGIANS OF SELF-DUAL NONLINEAR
ELECTRODYNAMICS}
\author{E.A. Ivanov, B.M. Zupnik\\\vspace{0.2cm}
Bogoliubov Laboratory of Theoretical Physics, JINR\\
Dubna, Moscow Region, 141980, Russia}

\begin{document}

\begin{abstract}
We elaborate on a new representation of Lagrangians of $4D$ nonlinear
electrodynamics including the Born-Infeld theory as a particular case.
In this new formulation, in parallel with the standard Maxwell field
strength $F_{\alpha\beta}, \bar{F}_{\dot\alpha\dot\beta}$, an auxiliary
bispinor field $V_{\alpha\beta}, \bar{V}_{\dot\alpha\dot\beta}$ is
introduced. The gauge field strength appears only in bilinear terms
of the full Lagrangian, while the interaction Lagrangian $E$ depends on
the auxiliary fields, $E = E(V^2, \bar V^2)$. The generic nonlinear
Lagrangian depending on $F,\bar{F}$ emerges as a result of eliminating
the auxiliary fields. Two types of self-duality inherent in the nonlinear
electrodynamics models admit a simple characterization in terms of the
function $E$. The continuous $SO(2)$ duality symmetry between nonlinear
equations of motion and Bianchi identities amounts to requiring $E$ to be
a function of the $SO(2)$ invariant quartic combination $V^2\bar V^2$,
which explicitly solves the well-known self-duality condition for
nonlinear Lagrangians. The discrete self-duality (or self-duality under
Legendre transformation) amounts to a weaker condition $E(V^2, \bar{V}^2)
= E(-V^2, -\bar{V}^2)$. We show how to generalize this approach to a
system of $n$ Abelian gauge fields exhibiting $U(n)$ duality. The
corresponding interaction Lagrangian should be $U(n)$ invariant function
of $n$ bispinor auxiliary fields. \vspace{1pc}
\end{abstract}

\maketitle
\section{Introduction}

It is well known that the on-shell $SO(2)$ ($U(1)$) duality invariance of
Maxwell equations can be generalized to the whole class of the nonlinear
electrodynamics models, including the famous Born-Infeld theory. The
condition of $SO(2)$ duality can be formulated as a nonlinear differential
equation for the Lagrangian of these theories \cite{GZ,GR,KT}. Up to now,
the general solution of this equation was analysed only in the framework
of proper power expansions. Here, based on our recent paper \cite{IZ}, we
present details of new representation for the Lagrangians of these
theories. Alongside with the electromagnetic field strength $F_\ab,
\bar{F}_{\dot\alpha\dot\beta}$, it involves an auxiliary tensor (bispinor)
field $V_{\alpha\beta}, \bar V_{\dot\alpha\dot\beta}$. It will be shown
that the $SO(2)$ duality condition can be explicitly solved in this
formalism, without resorting to any perturbative expansion. The general
Lagrangian solving this constraint is a sum of an interaction term
$E(V^2,\bar{V}^2)$ depending only on the $U(1)$-invariant combination of
the auxiliary fields, $E = \tilde{E}(V^2\bar V^2)$, and non-invariant
terms which are bilinear in $V$ and $F$. The Lagrangian involving only
the Maxwell field strengths emerges as a result of eliminating the tensor
auxiliary fields by their algebraic equations of motion. More general
nonlinear electrodynamics Lagrangians respecting the so-called discrete
self-duality (or duality under Legendre transformation) also admit a
simple characterization in terms of the function $E$. In this case it
should be even, $E(V^2,\bar V^2) = E(-V^2, -\bar V^2)$, and otherwise
arbitrary.

In Sect. 2 we give a brief account of the continuous and ``discrete''
dualities in nonlinear electrodynamics in the conventional approach. A
novel representation of the appropriate Lagrangians via bispinor
auxiliary fields $V_\ab$ and $\bar{V}_{\da\db}$ is discussed in Sect. 3.
Two examples of duality-invariant models in the new setting, including
the Born-Infeld theory, are presented in Sect. 4. An extension to $U(n)$
duality-invariant systems of $n$ Abelian gauge fields is given in Sect. 5.
The corresponding Lagrangian in the $V, F$ representation is fully
specified by the interaction term which is an $U(n)$ invariant function
of $n$ auxiliary fields $V_\ab^k$. The discrete self-duality also amounts
to a simple restriction on the interaction function.

\section{ Self-dualities in nonlinear\\ electrodynamics}

We shall discuss nonlinear 4D electrodynamics models which reveal duality
properties and include the free Maxwell theory and Born-Infeld theory as
particular cases. Detailed motivations why such models are of interest to
study can be found, e.g., in \cite{KT}.
\vspace{0.3cm}

\noindent{\it 2.1 Continuous on-shell $SO(2)$ duality.}
In the spinor notation, the Maxwell field strengths are defined by
\bea
&&F_\ab(A)\equiv {1\over2}(\partial_{\alpha}^\db A_{\beta\db}+
\partial_{\beta}^\db A_{\alpha\db})~,\nn\\
&&\bar{F}_{\da\db}(A)\equiv {1\over2}(\partial^\beta_{\da}A_{\beta\db}+
\partial^\beta_{\db}A_{\beta\da})~,
\lb{defF}
\eea
where $\padb={1\over2}(\sigma^m)_\adb\partial_m$ and $A_\adb$ is the
corresponding vector gauge potential. Below we shall sometimes treat
$F_\ab$ and $\bar{F}_{\da\db}$ as independent variables, without assuming
Eqs. \p{defF}.

Let us introduce the Lorentz-invariant complex variables
\bea
&& \varphi = F^\ab F_\ab~,
\quad \bar\varphi  =
\bar{F}^{\da\db}\bar{F}_{\da\db}~.
\lb{compl}
\eea
In this representation, two independent invariants which one can construct
out of the Maxwell field strength in the standard vector notation take the
following form:
\bea
&&F^{mn}F_{mn}=2( \vp+\bvp)~,\nn\\
&&F_{mn}
\widetilde{F}^{mn} =-2i(\vp-\bvp)~. \lb{vecsp}
\eea
Here
$$
F_{mn}=\partial_mA_n-\partial_nA_m~,\q\widetilde{F}^{mn}
={1\over2}\,\varepsilon^{mnpq}F_{pq}.
$$

It will be convenient to deal with dimensionless $F_{\alpha\beta},
\bar{F}_{\dot\alpha\dot\beta}$ and $\vp, \bvp$, introducing a coupling
constant $f$, $[f] = 2$. Then the generic nonlinear Lagrangian
${\cal L}(F,\bar F) = f^{-2}L(\vp, \bvp)$, where
\be
L(\vp,\bvp) =
-{1\over2}(\vp+\bvp)+L_{int}(\vp,\bvp)
\lb{nonl}
\ee
and $L_{int}(\vp, \bvp)$ collects all possible self-interaction terms of
higher-order in $\vp, \bvp$.

We shall use the following notation for the derivatives of the Lagrangian
$L(\vp,\bvp)$ \footnote{In these and some subsequent relations it is
assumed that the functional argument $F$ stands for both $F_\ab$ and
$\bar{F}_{\da\db}$; we hope that this short-hand notation will not give
rise to any confusion.}
\bea
&&P_\ab(F)\equiv i\partial L/\partial F^\ab=2iF_\ab
 L_\vp~,\lb{defP}\\
&&L_\vp= \partial L/\partial \vp~,\q L_\bvp= \partial L/ \partial
\bvp~,\;\nn \\
&& L_{\vp\bvp}= \partial^2 L/\partial \vp\partial\bvp\ldots~,\nn
\eea
and for the bilinear combinations of them
\bea
&&
\pi\equiv P^\ab P_\ab=-4\vp(L_\vp)^2~,\nn\\
&&\bar\pi \equiv
\bar{P}^{\da\db}\bar{P}_{\da\db}=-4\bvp(L_\bvp)^2~.\lb{Pvar}
\eea
In the vector notation, the same quantities read
\bea
&&\widetilde{P}^{mn}\equiv{1\over2}\,\varepsilon^{mnpq}P_{pq}=
2 \partial L/\partial F_{mn}\nn\\
&&{i\over2}P_{mn}\widetilde{P}^{mn}=\pi-\bar{\pi}~.\nn
\eea

The nonlinear equations of motion have the following form in this
representation:
\bea
&& {\cal E}_\ada(F)\equiv\partial_\alpha^\db \bar P_{\da\db}(F)
-\partial^\beta_\da P_\ab(F)= 0~. \lb{BIeq}
\eea
These equations, together with the Bianchi identities
\bea
&&{\cal B}_\ada(F)\equiv
\partial_\alpha^\db \bar F_{\da\db}-\partial^\beta_\da F_\ab
= 0~, \lb{Bian}
\eea
constitute a set of first-order equations in which one can treat $F_\ab$
and $\bar{F}_{\da\db}$ as {\it unconstrained} conjugated variables.

This set is said to be duality-invariant if the Lagrangian $L(\vp, \bvp)$
satisfies certain nonlinear condition \cite{GZ,GR,KT}. The precise form
of this self-duality condition in the spinor notation is
\bea
&S[F,P(F)]\equiv F^\ab F_\ab+P^\ab P_\ab-\mbox{c.c.}
\nn\\
&= \vp + \pi -\bvp - \bar\pi \nn \\
&=\vp - \bvp - 4\,[\vp(L_\vp)^2 - \bvp(L_\bvp)^2]=0.
&\lb{sdI}
\eea
Let us define the nonlinear transformation
\bea
&& \delta_\omega F_\ab=\omega\,P_\ab(F)\equiv 2i\,\omega\,F_\ab L_\vp~,
\lb{Ftrans}
\eea
where $\omega$ is a real parameter. This transformation is a nonlinear
realization of the $SO(2)$ group, provided the condition \p{sdI} is
satisfied. Indeed, in this case $F_{\alpha\beta}$ and $P_\ab(F)$ form
an $SO(2)$ vector
\bea
&&\delta_\omega P_\ab(F)=-4\,\omega\,F_\ab{\partial\over\partial\vp}
[\vp(L_\vp)^2-\bvp( L_\bvp)^2]\nn\\
&&=-\omega F_\ab~. \lb{Ptrans1}
\eea
The set of equations \p{BIeq}, \p{Bian} is clearly invariant under
these transformations
\bea
&&\delta_\omega \left(\begin{array}{c}
{\cal B}_\ada    \\
{\cal E}_\ada
\end{array}\right)=\left(\begin{array}{cc}
0  & \omega    \\
-\omega  & 0
\end{array}\right)\left(\begin{array}{c}
{\cal B}_\ada    \\
{\cal E}_\ada
\end{array}\right).
\eea
Thus these transformations are an obvious generalization of the $SO(2)$
duality transformation in the Maxwell theory:
\bea
&& \delta_\omega  F_\ab=-i\omega\, F_\ab~, \q \delta_\omega
\bar{F}_{\da\db}= i\omega\,\bar{F}_{\da\db}~,
\eea
which is a symmetry of the vacuum Maxwell equation
$\partial^\beta_\da F_\ab=0$.

The authors of \cite{GR} analyzed the self-duality condition \p{sdI} as an
equation for the unknown $L$, using the power expansion in some variable.
They have found that in each order of the perturbation theory the general
solution for $L$ is specified by some arbitrary function of the single
variable $\vp\bvp$. In Sect. 3 we propose the explicit solution to this 
problem beyond the power expansions, taking advantage of a new representation 
of self-dual Lagrangians via bispinor auxiliary fields.

It should be pointed out that the $SO(2)$ duality transformations in the
standard setting described above cannot be realized on the vector
potential $A_m$; they provide a symmetry between the equations of motion
and Bianchi identity and as such define {\it on-shell} symmetry. The
manifestly $SO(2)$ duality-invariant Lagrangians can be constructed in the
formalism with additional vector and auxiliary fields \cite{PST}. Here we
shall not discuss connections with this extended formalism.

Although the Lagrangian $L(\vp,\bvp)$ satisfying \p{sdI} is not invariant
with respect to transformation \p{Ftrans}, one can still construct,
out of $\vp$ and $\bvp$, the $SO(2)$ invariant function
\bea
&&I[F,P(F)]\equiv L+{i\over2}(FP-\bar{F}\bar{P})\nn\\
&&=L-\vp L_\vp-\bvp L_\bvp\equiv I(\vp,\bvp)~,\lb{invar}\\
&&\delta_\omega I(\vp,\bvp)={i\over2}\,\omega\,(\vp+\pi-\bvp-\bar{\pi})=0
~,\nn
\eea
where $FP-\bar{F}\bar{P}=P_\ab F^\ab-\bar{P}_{\da\db}\bar{F}^{\da\db}$.
This function cannot be taken as a Lagrangian since it contains
no free part.
\vspace{0.3cm}

\noindent{\it 2.2 Self-duality under Legendre transformation.}
For what follows we shall need a first-order representation of the action
corresponding to the Lagrangian \p{nonl}, such that the Bianchi identities
\p{Bian} are implemented in the action with the appropriate Lagrange
multipliers and so $F_\ab, \bar F_{\da\db}$ are unconstrained complex
variables. This form of the action is given by
\bea
&&{1\over f^2}\int d^4x\widetilde{L}[F,F^\D(B)]
={1\over f^2}\int d^4x\{L(\vp,\bvp)\nn\\
&&+i[FF^\D(B)-\bar{F}\bar{F}^\D(B)]\}~, \lb{1stor}
\eea
where
\bea
&& F^\D_{\alpha\beta}(B)\equiv{1\over2}(\partial_{\alpha}^\db B_{\beta\db}
+\partial_{\beta}^\db B_{\alpha\db})~.\lb{PB}
\eea
Varying with respect to the Lagrange multiplier $B_\adb$, one obtains just
the Bianchi identities for $F_\ab, \bar F_{\da\db}$ \p{Bian}. Solving them
in terms of the gauge potential $A_{\alpha\dot\beta}$ and substituting the
result into \p{1stor}, we come back to \p{nonl}. On the other hand, the
multiplier  $B_\adb$ is defined up to the standard Abelian gauge
transformation, which suggests interpreting $B_\adb$ and
$F^\D_{\alpha\beta}(B)$ as the {\it dual} gauge potential and gauge field
strength, respectively. Using the algebraic equations of motion for the
variables $F_\ab, \bar F_{\da\db}$, one can express the action \p{1stor}
in terms of $F^\D_{\alpha\beta}(B), \bar F^\D_{\da\db}(B)$. If the
resulting action has the same form as the original one in terms of
$F_\ab(A), \bar F_{\da\db}(A)$, the corresponding electrodynamics model
is said to enjoy the ``discrete'' self-duality. This sort of duality
should not be confused with the on-shell continuous $SO(2)$ duality
discussed earlier. The relevant $SO(2)$ symmetry is realized on the
variables $F_\ab$ according to \p{Ftrans} and it is not defined on
the dual vector potential $B_\adb$. However, as we shall see soon, any
$L(\vp, \bvp)$ solving the constraint \p{sdI} corresponds to a system
revealing the discrete self-duality. The inverse statement is not
generally true, so the class of nonlinear electrodynamics actions
admitting $SO(2)$ duality of equations of motion forms a subclass in the
variety of actions which are self-dual in the ``discrete'' sense.

Let us elaborate on this in some detail. The dual picture is achieved by
varying \p{1stor} with respect to the independent variables $F_\ab,
\bar{F}_{\da\db}$, which yields
\be
F^\D_{\alpha\beta}(B)=i \partial L/\partial F^\ab \equiv P_\ab(F) =
2iF_\ab L_\vp~,
\label{FDF}
\ee
where $P_\ab(F)$ is the same as in \p{defP}. Substituting the solution of
this algebraic equation, $F_\ab = F_\ab(F^\D)$, into \p{1stor} gives us
the dual Lagrangian $\hat{L}(F^D)$
\bea
&& \hat{L}(F^\D)= \hat{L}(\pi^\D, \bar\pi^\D) \equiv
\widetilde{L}[F(F^\D), F^\D]~,
\lb{hatL}
\eea
where $\pi^\D \equiv F^{\D\,\ab}F^\D_{\ab} = \pi(F)$ and $\pi, \bar\pi$
were defined in \p{Pvar}.

Using \p{FDF} and its conjugate, as well as the definitions \p{1stor},
\p{hatL}, one can explicitly check the property
\bea
&& F_\ab=-i \partial \hat{L}/\partial F^{\D\,\ab} \quad
\mbox{and c.c.}~.
\eea
Due to this relation, and keeping in mind the inverse one \p{FDF},
one can treat the equation
\bea
&&\hat{L}(\pi,\bar\pi)
=L(\vp,\bvp)+i(FP-\bar{F}\bar{P})\lb{stleg}
\eea
as setting the direct and inverse Legendre transforms
$L\leftrightarrow \widetilde{L}$ between two functions of 6 complex
variables
\bea
&&F_\ab~\Rightarrow~P_\ab=i\partial L/\partial F^\ab~,\nn\\
&&dL=-iP_\ab dF^\ab+i\bar{P}_{\da\db}d\bar{F}^{\da\db}~,\lb{dirleg}\\
&&P_\ab~\Rightarrow~F_\ab=-i \partial \hat{L}/\partial P^\ab~,
\nn\\
&&d\hat{L}=iF^\ab dP_\ab-i\bar{F}^{\da\db}d\bar{P}_{\da\db}~.
\lb{invleg}
\eea

Within this interpretation, the ``discrete'' self-duality defined
above and amounting to the condition
\bea
&& \hat{L}(\pi, \bar\pi) = L(\pi, \bar\pi)~, \label{discr}
\eea
can be equivalently called ``self-duality under Legendre transformation''
\cite{GZ,KT}.

Let us show  that the $SO(2)$ duality condition \p{sdI} indeed guarantees
the self-duality under Legendre transformation \p{discr}. The simplest
proof of this statement (see, e.g., \cite{KT}) makes use of the finite
discrete $SO(2)$ transformation
\bea
&& F_\ab\rightarrow P_\ab~,\q P_\ab\rightarrow -F_\ab \lb{discrSO}
\eea
and the invariance of function \p{invar} under the global version
of the $SO(2)$ transformations \p{Ftrans}. Due to the latter property,
\p{invar} is invariant with respect to \p{discrSO} too
$$
L(\vp,\bvp)+{i\over2}PF-{i\over2}\bar{P}\bar{F}=
L(\pi,\bar\pi)-{i\over2}FP+{i\over2}\bar{F}\bar{P}.
$$
Comparing this relation with \p{stleg}, we arrive at the condition
\p{discr}.

For the dual Lagrangian $\hat L(\pi, \bar\pi)$ one can construct
the 1st order action similar to \p{1stor}
\bea
&f^{-2}\int d^4x \widetilde{L}[P, F(A)]=f^{-2}\int d^4x
[\hat L(\pi,\bar{\pi})&\nn\\
&-iPF(A)+i\bar{P}\bar{F}(A)].&\lb{stand}
\eea
where $F_\ab(A), \bar F_{\da\db}(A)$ were defined in \p{defF}. The
self-dual case (in the ``discrete'' sense) corresponds to identifying
$\hat L(\pi,\bar{\pi})= L(\pi,\bar{\pi})$. Varying \p{stand} with respect
to the gauge potential $A_{\alpha\dot\beta}$ produces Bianchi identities
for the originally unconstrained variable $P_\ab, \bar P_{\da\db}$ as the
corresponding equations of motion and so implies
$$
P_\ab = F^D_\ab(B)~, \q \bar P_{\da\db} = \bar F^D_{\da\db}(B)~.
$$
On the other hand, the equation of motion for $P^\ab$ in this
representation yields
\bea
&& F_\ab =-i\partial L/\partial P^\ab = -2iP_\ab L_\pi~.
\lb{Peq}
\eea
Solving this equation for the unknown $P_\ab$ as a function of $F_\ab(A),
\bar{F}_{\da\db}(A)$, we come back to the action corresponding to the
original Lagrangian \p{nonl}.

The on-shell $SO(2)$ duality formulated earlier in terms of the
variables $F_\ab, P_\ab(F)$ admits an equivalent formulation in
terms of the dual variables $P_\ab, F_\ab(P)$:
\bea
&& \delta_\omega P_\ab= - \omega\,F_\ab(P) = 2i \omega\,P_\ab L_\pi
~.\lb{Ptrans}
\eea
The condition of $SO(2)$ self-duality has the following form in this
representation:
\bea
&& S[F(P), P]=\pi[1-4 (L_\pi)^2]-\mbox{c.c.}=0~.
\eea
The function $S [F(P), P]$ is obtained by substituting $F_\ab(P)$
from \p{Peq} into the universal bilinear form \p{sdI}.

\section{A new form of the actions of nonlinear electrodynamics  and
self-dualities}
\noindent{\it 3.1 Nonlinear electrodynamics Lagrangians in a new setting.}
The recently constructed $N=3$ supersymmetric extension of the Born-Infeld
theory \cite{IZ} suggests a new representation for the actions of
nonlinear electrodynamics discussed in the previous Section.

The infinite-dimensional off-shell $N=3$ vector multiplet contains
gauge field strengths \p{defF}  and auxiliary fields $H_\ab$ and
$\bar{H}_{\da\db}$ \cite{HS}. The gauge field part of the off-shell
super $N=3$ Maxwell component Lagrangian is \footnote{In the rest of the
paper we put the overall coupling constant $f$ equal to 1.}
\bea
&& {1\over16}[\,h+\bar{h} -6\,(\bar{H}\bar{F}+ H
F)+ \vp+\bvp  \,]~, \lb{LFH}
\eea
where $h=H^\ab H_\ab$ and $HF=H^\ab F_\ab(A)$.
Eliminating
the auxiliary fields $H_{\alpha\beta}, \bar H_{\da\db}$ by their algebraic
equations of motion we arrive at the standard Maxwell action
\bea
&& L_2(F) = - {1\over2}(\vp + \bvp)~. \lb{maxw1}
\eea

The $N=3$ off-shell superfield strengths contain the following
combinations of fields \cite{IZ}:
\bea
&&V_{\alpha\beta} = {1\over 4}\, (
H_{\alpha\beta} + F_{\alpha\beta})~, \nn\\
&&\bar V_{\da\db} = {1\over 4}\,(
\bar H_{\da\db} + \bar F_{\da\db})~. \lb{redef}
\eea

The free Maxwell Lagrangian \p{LFH}, being rewritten
through $V_{\alpha\beta}, \bar V_{\da\db}$, reads
\be
\cL_2(V,F)=\nu+ \bnu- 2\,(V F+\bar{V}\bar{F})
+{1\over2}(\vp+\bvp)~, \lb{auxfree}
\ee
where
\bea
&&\nu\equiv V^\ab V_\ab~,\q\bnu\equiv\bar{V}^{\da\db}\bar{V}_{\da\db}~,
\nn\\
&& VF\equiv V^\ab F_\ab~,\q\bar{V}\bar{F}\equiv \bar{V}^{\da\db}
\bar{F}_{\da\db}.
\eea
Eliminating $V^\ab$ by its algebraic equation of motion,
\bea
&& V^\ab = F^\ab~, \q \bar V^{\da\db} = \bar F^{\da\db}~,\lb{free1}
\eea
we arrive at the free Lagrangian \p{maxw1}.

Our aim will be to find a nonlinear extension of the free Maxwell
Lagrangian using the $N=3$ supersymmetry-inspired form $\cL_2(V, F)$ of
it, eq. \p{auxfree}, such that this extension becomes the generic
nonlinear Lagrangian $L(F^2,\bar{F}^2)$, eq. \p{nonl}, after eliminating
the auxiliary fields $V_{\alpha\beta}, \bar{V}_{\dot\alpha\dot\beta}$ by
their algebraic ({\it nonlinear}) equations of motion.

By Lorentz covariance, such a nonlinear Lagrangian has the following
general form:
\bea
&& \cL[V,F(A)] = \cL_2[V,F(A)] + E(\nu,\bnu)~, \lb{legact}
\eea
where $E$ is a real function encoding self-interaction. Varying the action
with respect to $V_\ab$, we derive the algebraic relation between $V$ and
$F(A)$ in this formalism
\bea
&& F_\ab(A) = V_\ab(1+ E_\nu) \quad \mbox{and c.c.}~, \lb{FV}
\eea
where $E_\nu\equiv\partial E(\nu,\bnu)/\partial\nu$. This relation is a
generalization of the free equation \p{free1} and it can be used to
eliminate the auxiliary variable $V^\ab$ in terms of
$F^\ab$ and $\bar F^{\da\db}$, $V_\ab~\Rightarrow~V_\ab[F(A)]$
(see eq. \p{VF} below). The natural restrictions on the interaction
function $E(\nu, \bnu)$ are
\bea
&& E(0,0)=0~,\q E_\nu(0,0)= E_\bnu(0,0) = 0~,
\eea
which mean that its $(\nu, \bnu)$-expansion does not contain constant and
linear terms. Clearly, given some non-singular interaction Lagrangian
$L_{int}(\vp, \bvp)$ in \p{nonl}, one can pick up the appropriate
function $E(\nu, \bnu)$, such that the elimination of $V^\ab,
\bar V^{\da\db}$ by \p{FV} yields just this self-interaction.
Thus \p{legact} with an arbitrary (non-singular) interaction function $E$
is another form of generic nonlinear electrodynamics Lagrangian \p{nonl}.
The second equation of motion in this representation, obtained by
varying \p{legact} with respect to $A_\ada$, has the form
\bea
&& \partial^\beta_\da[F_\ab(A)-2V_\ab]+\mbox{c.c.}=0~.\lb{FV3}
\eea
After substituting $V_\ab = V_\ab[F(A)]$ from \p{FV}, eq. \p{FV3} becomes
the dynamical equation for $F_\ab(A), \bar F_{\da\db}(A)$ corresponding
to the generic Lagrangian  \p{nonl}. Comparing \p{FV3} with \p{BIeq}
yields the relation
\bea
&& P_\ab(F) = i\left[ F_\ab - 2V_\ab(F) \right]~, \lb{imprel}
\eea
where $P_\ab(F)$ was defined in \p{defP}.

Let us elaborate in more detail on the relation of the $V, F$
representation of the nonlinear electrodynamics Lagrangians to the
original ``minimal'' one \p{nonl} which involves only $F_\ab$ and
$\bar F_{\da\db}$. The general solution of the algebraic equation \p{FV}
for $V_\ab$ can be written as
\bea
&& V_\ab(F)=F_\ab G(\vp,\bvp)\lb{VF}~.
\eea
The transition function $ G(\vp, \bvp)$ can be found from the basic
requirement that  \p{legact} coincides with the initial nonlinear
action after eliminating $V_\ab, \bar V_{\dot\alpha\dot\beta}$
\bea
&& \cL[V(F),F]=L(\vp,\bvp)~.\lb{EL1}
\eea
Using eq.\p{VF}, one can obtain the relations
\bea
&&\nu=\vp G^2~,\q\bnu=\bvp\bar{G}^2\lb{nuvp}~, \\
&&V(F)F=\vp G~,\q\bar{V}(F)\bar F=\bvp\bar{G}~.\lb{VF2}
\eea
After substituting these expressions in \p{EL1} with making use of the
explicit expressions \p{auxfree}, \p{legact}, eq. \p{EL1} can be
rewritten as
\bea
&&E(\nu,\bnu) =
L(\vp, \bvp)-{1\over2}(\vp+\bvp) \nn \\
&& +\,\vp G(2-G) +\bvp\bar{G}(2-\bar{G})~.
\lb{base2}
\eea
One should also add the relations:
\bea
&& G^{-1}=1+E_\nu~,\q\bar{G}^{-1}=1+E_\bnu~, \lb{GE}
\eea
which follow from comparing \p{VF} with \p{FV}.

Differentiating \p{base2} with respect to $\vp$ and using
the relations
$$
\frac{\partial \nu}{\partial \vp} = G^2 +2\vp G\frac{\partial G}
{\partial\vp}~, \quad
\frac{\partial\bnu}{\partial\vp} = 2\bvp\bar G\frac{\partial \bar G}
{\partial\vp}~,
$$
one obtains the simple expression for transition functions
\bea
&& G(\vp, \bvp)= {1\over 2}-L_\vp~.
\lb{exprG}
\eea
A useful corollary of this formula and eqs. \p{nuvp}, \p{GE} is the
relation
\bea
&& \nu E_\nu={1\over4}\vp(1-4L^2_\vp)~. \lb{corol}
\eea

Given a fixed $L(\vp, \bvp)$, one can express $\vp,\bvp$ in terms of
$\nu, \bnu$ using eqs. \p{nuvp}, \p{exprG} and then restore the explicit
form of $E(\nu, \bnu)$ by \p{base2}. Conversely, given $E(\nu, \bnu)$,
one can restore $L(\vp, \bvp)$. In practice, finding out such explicit
relations is a rather complicated task, as we shall see on two examples
in Sect. 4.
\vspace{0.3cm}

\noindent{\it 3.2 Self-dualities revisited.}
Until now we did not touch any issues related to self-dualities. A link
with the consideration in the previous Section is established by Eq.
\p{imprel} which relates the functions $P_\ab(F)$ and $V_\ab(F)$.

Substituting this into the $SO(2)$ duality condition \p{sdI} and
making use of eq. \p{corol} we find
\bea
&&S[F,P(F)]=[F^\ab - V^\ab(F)]V_\ab(F)
-\mbox{c.c}\nn\\ &&=\nu E_\nu-\bnu E_\bnu~. \lb{Vsd}
\eea
Thus passing to the $V, F$ representation allows one to reduce the
nonlinear differential equation \p{sdI} to a {\it linear} differential
equation for the function $E(\nu, \bnu)$:
\be
\mbox{Eq.\p{sdI}} \quad \Leftrightarrow \quad\nu E_\nu-\bnu E_\bnu = 0~.
\lb{newsd}
\ee
The corresponding realization of the $SO(2)$ (or $U(1)$) transformations
\p{Ftrans}, \p{Ptrans1} in terms of $F^\ab$ and $V^\ab(F)$ is given by
\bea
&&\delta_\omega V_\ab(F)=-i \omega V_\ab(F)~,\lb{Vtrans}\\
&&\delta_\omega F_\ab =i \omega [F_\ab -2V_\ab(F)]~.\lb{Ftrans1}
\eea

It is worth recalling that the on-shell $SO(2)$ duality transformation
mixes the dynamical equations of motion for $F_\ab, \bar F_{\da\db}$
with the Bianchi identities \p{Bian}, and so in this extended set of
equations one should treat $F_\ab, \bar F_{\da\db}$ as {\it unconstrained}
conjugated complex variables (without explicitly solving \p{Bian} in terms
of gauge potential $A_{\alpha\db}$). Correspondingly, the algebraic
relation \p{FV} should be viewed to connect two independent sets of
variables, $(F_\ab, \bar F_{\da\db})$ and $(V_\ab, \bar{V}_{\da\db})$:
\bea
&& F_\ab=V_\ab(1+ E_\nu)~.\lb{FV2}
\eea
Due to this relation, one can equivalently formulate the dynamics and
duality transformations in terms of either the set $(F_\ab,
\bar{F}_{\da\db})$ or the set $(V_\ab, \bar V_{\da\db})$. Respectively,
one can choose as the basic one either the transformation \p{Ftrans1} or
\p{Vtrans}. It is important to be sure that \p{FV2} is covariant under
the transformations \p{Vtrans},\p{Ftrans1}. It is straightforward to
check that both \p{FV2} and the constraint \p{newsd} itself are indeed
covariant under these transformations on the surface of eq. \p{newsd}.

It is important to emphasize that the new form \p{newsd} of the
self-duality constraint \p{sdI} admits a transparent interpretation as
the condition of invariance of $E(\nu,\bnu)$ with respect to the
$U(1)$ transformations \p{Vtrans}
\bea
&& \delta_\omega E = 2i\omega (\bnu E_\bnu - \nu E_\nu) = 0~.
\eea

The general solution of \p{newsd} is a function $\tilde{E}(a)$ which
depends on the single real $U(1)$ invariant variable $a=\nu\bnu$ quartic
in the  auxiliary fields $V_\ab$ and $\bar{V}_{\da\db}$
\bea
&& E_{sd}(\nu, \bnu) = \tilde{E}(a) = \tilde{E}(\nu\bnu)~,\;
\tilde{E}(0)=0~.
\label{condfin}
\eea
Thus we come to the notable result that the {\it whole} class of nonlinear
extensions of the Maxwell action admitting the on-shell $SO(2)$ duality
is parametrized by an arbitrary $SO(2)$ invariant real function of
one argument $E_{sd} = \tilde{E}(\nu\bnu)$ in the representation
\p{legact}. The remarkable property of $E_{sd}$ is that only terms
$\sim \nu^n\bnu^n$ can appear in its power expansion. Below we shall
present this expansion for two examples, including the most interesting
case of Born-Infeld theory.

In the $V, F$ representation it is very easy to construct invariants of
the $SO(2)$ duality rotations. Besides the function $E(\nu, \bnu)$ itself,
one more real invariant combination of $V_\ab$ and $F_\ab(V)$ is as
follows
\bea
&&I_0(V,F)=\nu+\bnu-V F-\bar{V}\bar{F}\nn\\
&& =-\nu E_\nu-\bnu E_\bnu~.\lb{Vinv1}
\eea

Finally, let us examine which restrictions on the interaction Lagrangian
$E(\nu, \bnu)$ are imposed by the requirement of the ``discrete''
self-duality with respect to the exchange $F(A) \leftrightarrow F^D(B)$.
For this we shall need a first-order representation of the Lagrangian
\p{legact} similar to \p{1stor}.

Let us treat $\cL(V,F)$ in eq.\p{legact} as a function of two independent
variables and implement the Bianchi identities for $F_\ab,
\bar{F}_{\da\db}$ (amounting to the expressions \p{defF}) in the
Lagrangian via the dual field-strength $F^D_\ab(B)$ \p{PB}:
\bea
&& \widetilde{\cL}[V,F,F^\D(B)] \equiv \cL(V,F) \nn \\
&& +\,i[F^\D(B)F - \bar{F}^\D(B)\bar{F}]~. \lb{modB}
\eea
The algebraic $V^\ab$ equation of motion $\partial \widetilde{L}
/\partial V^\ab = 0$ is just the relation \p{FV2}. On the other hand,
varying \p{modB} with respect to $F_\ab$, one obtains the linear relation
\bea
&& F_\ab-2V_\ab=-iF^\D_\ab(B)  \q \mbox{and c.c.} \lb{FVP}
\eea
as the corresponding equation of motion. The Bianchi identities for
$F^D_\ab(B)$ following from the definition \p{PB} imply for $F_\ab =
F_\ab(A)$ just the dynamical equation \p{FV3} (with $V_\ab$ expressed in
terms of $F_\ab, \bar F_{\da\db}$ from \p{FV2}). This proves the
equivalence of the dynamics described by \p{modB} with that associated
with \p{legact} or \p{nonl}.

The function $\widetilde{\cL}[V,F,F^D(B)]$ in \p{modB} contains only
quadratic and linear terms in $F$ and $\bar{F}$, so one can explicitly
find the dual form of \p{modB} in terms of $F^D_\ab(B),
\bar{F}^D_{\da\db}(B)$ and $V_\ab, \bar V_{\da\db}$, expressing $F_\ab$
and $\bar F_{\da\db}$ from Eq. \p{FVP}:
\bea
&& \widetilde{\cL}[V,F(V,F^D), F^D]\equiv \widetilde{\cL}(U,F^D) \nn \\
&& = \cL_2(U,F^D)+ E(-u,-\bar{u})~, \label{mediate}
\eea
where
$$ U_\ab \equiv -iV_\ab~,\q u=U^\ab U_\ab~.
$$
Comparing the dual Lagrangian \p{mediate} with the original one
\p{legact}, we observe that the necessary and sufficient condition of the
discrete self-duality is the following simple restriction on the function
$E$ \cite{IZ}
\bea
&& E(\nu,\bnu)=E(-\nu,-\bnu)~. \lb{discr1}
\eea
Obviously, an arbitrary $SO(2)$-invariant function $E_{sd} =
\tilde{E}(\nu\bnu)$ corresponding to a $SO(2)$ self-dual system
automatically satisfies the discrete self-duality condition \p{discr1}.
This elementary consideration provides us with a simple proof of the fact
(mentioned in Sect. 2) that the $SO(2)$ self-dual systems constitute a
subclass in the set of those enjoying the discrete self-duality.

\section{Examples of self-dual systems}

\noindent{\it 4.1 Born-Infeld theory.}
The Lagrangian of the Born-Infeld theory has the following form
in terms of complex invariants \p{compl}
\bea
&& L_{\B\I}(\vp,\bvp) =
\left[1-Q(\vp,\bvp)\right]~, \lb{biact}
\eea
where
\bea
&&Q(\vp,\bvp) = \sqrt{1+X}~,\nn\\
&& X(\vp , \bvp) \equiv
(\vp+\bvp)+(1/4)(\vp-\bvp)^2~.\lb{not}
\eea
The power expansion of the BI-lagrangian is
\bea
&L_{\B\I}=-{1\over2}(\vp+\bvp)+{1\over2}\vp\bvp
-{1\over2^2}\vp\bvp(\vp+\bvp)&\nn\\
&+{1\over2^3}\vp\bvp(3\vp\bvp+
\vp^2+\bvp^2)+O(\vp^5)~.&\lb{BIpert}
\eea

In the BI theory the function  \p{defP} has the following explicit form
\bea
&&P_\ab(F)=i\frac{\partial L_{\B\I}}{\partial F^\ab}\nn\\
&&=-iF_\ab Q^{-1}(\vp,\bvp)[1+{1\over2}(\vp-\bvp)]~.
\eea
It is easy to check that this function satisfies the $SO(2)$ self-duality
condition \p{sdI}, so BI theory belongs to the class of self-dual models
\cite{GZ,GR}.

Let us study the Lagrangian $V, F$ function $\cL_{\B\I}(V,F)$ \p{legact}
for this particular case. Our basic purpose will be to find the
corresponding function $E_{\B\I}(\nu, \bnu)$.

The function $G(\vp, \bvp)$ relating the variables $V_\ab$ and $F_\ab$
and defined by eq. \p{exprG}, is given by the expression
\bea
&&G \equiv g ={1\over 2}\left\{1 + Q^{-1}
\left [1 + {1\over2}(\vp - \bvp)\right]\right\} \lb{Gbi}\\
&& =1-{1\over2}\bvp+{1\over2}\vp\bvp+{1\over4}(\bvp)^2+\ldots~.\nonumber
\eea
It is easy to find the inverse relation
\bea
&& \vp=2\,\bar g\, \frac{1-\bar g }{[1 -(g +\bar g)]^2}~.\lb{zG}
\eea

Our aim is to find $E_{\B\I}$ as a function of the variables $\nu = V^2,
\bnu = \bar V^2$. As the first step, one expresses $\nu, \bnu$
in terms of $g$ and $\bar g$, using \p{VF2} and \p{zG}
\bea
&& \nu= \vp\, g^2 = 2\,\bar{g} g^2 \frac{1 -\bar g}{[1 -(g +\bar g)]^2}~.
\lb{xG}
\eea
Introducing
\bea
&& t \equiv \frac{g\bar g}{1 -(g+\bar
g)}~, \lb{deft}
\eea
one finds that $t$, as a consequence of
\p{xG} and the fact that $g(\vp =0) = 1$,
satisfies the following quartic equation:
\be
 t^4 + t^3-{1\over4} \nu\bnu = 0~, \; t(\nu = \vp =0) = -1~. \lb{quart}
\ee
It allows one to express $t$ in terms of $a \equiv \nu\bnu$
\be
 t(a)=-1-\frac{a}{4}+\frac{3a^2}{16}-\frac{15a^3}{64}+
\ldots~.
\ee
One can write a closed expression for $t(a)$ as the proper solution
of \p{quart}, but we do not present it here in view of its complexity.

Now we are ready to find $E_{\B\I}(\nu, \bnu)$. Taking into account the
explicit expressions \p{zG} and \p{xG} and substituting all this into
\p{base2}, one finally finds a simple expression for $E_{\B\I}(\nu,\bnu)$
through the real variable $t(a)$ \cite{IZ}
\bea
&&E_{\B\I}(a)=2[2t^2(a)+3t(a)+1]\nn\\
&&=
\frac{a}{2}-\frac{a^2}{8}+ \frac{3a^3}{32}+\ldots~.
\lb{explE}
\eea
\vspace{0.3cm}

\noindent{\it 4.2 One more example.}
Let us now consider the self-dual system corresponding to
the simplest choice of the function $E$ in the action \p{legact}
\bea
&& \check{E}={1\over2}\nu\bnu~\q (\nu \check{E}_\nu=\check{E})\lb{poly}
\eea
which is the lowest order self-dual approximation of $E_{\B\I}$.
This model is distinguished in that the  relation \p{FV2}
and the corresponding representation of equations of motion via
 variables $V$ are polynomial.

Using Eq.\p{GE} one can obtain the relation between variables $\nu, \bnu$
and $\vp, \bvp$ in this case
\bea
&& \vp=\nu(1+{1\over2}\bnu)^2 \q \mbox{and c.c.}~.\lb{vpnu}
\eea
Like in the previous example, we present here first terms in the power
expansion of the solution
\bea
&& \nu=\vp-\vp\bvp +\vp^2\bvp+{3\over4}\vp\bvp^2+\ldots~. \lb{1}
\eea
The corresponding expansion of the transition function is
\bea
&&\check{G}={1\over2}-\check{L}_\vp=(1+{1\over2}\bnu)^{-1}\nn\\
&&=1-{1\over2}\bvp+{1\over2}\vp\bvp+{1\over4}(\bvp)^2+\ldots \lb{2}
\eea
and one can directly find $\check{L}(\vp, \bvp)$ for this case. The
Lagrangian $\check{L}(\vp, \bvp)$ turns out to be highly non-polynomial,
despite the fact that the interaction in the $V, F$ representation is
specified by the simple monomial \p{poly}.

Note that the first two orders of the solution \p{2} coincide with the
corresponding terms in the Born-Infeld theory \p{Gbi}, while the terms
starting from the 3-rd order are different. It can be also checked that
the first three terms in the non-polynomial Lagrangian of this model
$\check{L}(\vp, \bvp)$ coincide with those in $L_{\B\I}$ \p{BIpert}.

\section{$U(n)$ self-duality}

Let us consider $n$ Abelian field-strengths
\bea
&& F^i_\ab~,\q \bar{F}^i_{\da\db}~,
\eea
where  $i=1,2\ldots n$. As the first step, one can realize the group
$SO(n)$ on these variables
\bea
&& \delta_\omega F^i_\ab=\xi^{ik}F^k_\ab~,\q \xi^{ki}=-\xi^{ik}~.\lb{On}
\eea
This group is assumed to define an off-shell symmetry of the corresponding nonlinear
Lagrangian $L(F^k, \bar F^k) = -{1\over 2}(F^iF^i)
-{1\over 2}(\bar{F}^i\bar{F}^i) + L_{int}(F^k, \bar{F}^k)$.

The $U(n)$ self-duality conditions for the Lagrangian $L(F^k,\bar{F}^k)$
generalizing the $U(1)$ condition \p{sdI} have been analyzed
in Refs.\cite{AT,ABMZ,KT}. In the spinor notation, these conditions read
\bea
&\cA^{[kl]}=(F^kP^l)-(F^lP^k)-\mbox{c.c.}
=0~,&\lb{Acond}\\
&\cS^{(kl)}=(F^kF^l)+(P^kP^l)-\mbox{c.c.}
=0~,&\lb{Scond}
\eea
where
\bea
&&P^k_\ab(F)\equiv i{\partial L\over\partial F^{k\ab}}
\\
&&(F^kP^l)\equiv F^{k\ab}P^l_\ab~, \q \mbox{etc}~. \nn
\eea

The condition \p{Acond} amounts to the $SO(n)$ invariance of the
Lagrangian and holds off shell. The second condition is the true analog
of \p{sdI}. It guarantees the covariance of the equations of motion for
$F^{k}_\ab,  \bar F^{k}_{\da\db}$ together with Bianchi identities,
\bea
&&{\cal E}^k_\ada(F)\equiv\partial_\alpha^\db \bar P^k_{\da\db}(F)
-\partial^\beta_\da P^k_\ab(F)
= 0~, \lb{Uneq}\\
&&{\cal B}^k_\ada(F)\equiv
\partial_\alpha^\db \bar F^k_{\da\db}-\partial^\beta_\da F^k_\ab
= 0~, \nn
\eea
under the  following nonlinear transformations:
\bea
&&\delta_\eta F_\ab^k=-\eta^{kl}P^l_\ab(F)~,\lb{Btran}\\
&&\delta_\eta P_\ab^k(F)=\eta^{kl}F^l_\ab\nn
\eea
where $\eta^{kl}=\eta^{lk}$ are real parameters. On the surface of the
condition \p{Scond} these transformations, together with \p{On}, form
the group $U(n)$. The $U(n)$ group structure becomes manifest
after passing to the new variables:
\bea
&&V^k_\ab(F)\equiv {1\over2}[F^k_\ab+iP^k_\ab(F)]~,\nn\\
&&\delta V^k_\ab(F)=\omega^{kl}V^l_\ab(F)~,\lb{Htran}\\
&&\omega^{kl}=\xi^{kl}+i\eta^{kl}~,\q \bar\omega^{lk}=-\omega^{kl}~.\nn
\eea

The particular solution of the $U(n)$ self-duality conditions \p{Acond},
\p{Scond} constructed so far \cite{ABMZ} is formulated in terms of the
algebraic equation for auxiliary scalar variables
$\chi^{kl}=\chi^{lk}$
\bea
&&L=-{1\over2}(\chi^{kk}+\bar\chi^{kk})~,\nn\\
&&\chi^{kl}+{1\over2}\chi^{kj}\bar\chi^{jl}=(F^kF^l)~.\lb{Unrep}
\eea
It generalizes a similar representation for the BI Lagrangian (or its
$N=1$ extension) \cite{BG,RT}
\bea
&&L_{\B\I}=-{1\over2}(\chi+\bar\chi)~,\nn\\
&&\chi+{1\over2}\chi\bar\chi=\vp=(FF)~,\lb{birep}
\eea
where $\chi$ is an auxiliary scalar. This $n=1$ version of Eq.\p{Unrep}
can be readily solved
$$\chi={1\over2}(\vp-\bvp)-L_{\B\I}~.$$
The solution for an arbitrary $n$ has been constructed in \cite{ABMZ}
within a perturbative expansion.

Passing to an analog of the $V, F$ representation in the $U(n)$ case
will allow us to find the {\it general} solution to \p{Acond}, \p{Scond}.

Let us define a new  representation for the $SO(n)$ invariant nonlinear
electrodynamics Lagrangians in terms of the Abelian gauge field strengths  $F^k_\ab(A^k)$
and auxiliary fields $V^k_\ab$
\bea
&& \cL(V^k,F^k)=(V^kV^k)+(\bar{V}^k\bar{V}^k) \nn \\
&& -\, 2(V^kF^k)-2(\bar{V}^k\bar{F}^k) +{1\over2}(F^kF^k) \nn \\
&& +\,{1\over2}(\bar{F}^k\bar{F}^k) + E(V^k,\bar{V}^k)~. \lb{UVF}
\eea
The real Lagrangian of interaction $E(V, \bar V)$ is $SO(n)$ invariant
by definition. In the case without interaction ($E=0$), the bilinear
part of \p{UVF} gives the standard free Lagrangian  of $n$ Abelian
fields,
\bea
&& L_2(F^k,\bar{F}^k)=-{1\over2}(F^kF^k)-{1\over2}(\bar{F}^k\bar{F}^k)~,
\eea
as the result of eliminating the auxiliary fields.

In the general case of $E\neq 0$ the algebraic equation for $V_\ab$ is
\bea
&& F^k_\ab=V^k_\ab+{1\over2}\frac{\partial E}{\partial V_k^\ab}
~.
\lb{FVrel}
\eea
Using the relation
\bea
 && P^k_\ab=i[F^k_\ab-2V^k_\ab]~,
\eea
one can rewrite the $U(n)$ self-duality conditions \p{Acond} and
\p{Scond} in this representation as follows
\bea
&& i(F^lV^k)-i(F^kV^l)-\mbox{c.c.}=0~, \nn \\
&& (F^lV^k)+(F^kV^l)-2(V^kV^l)-\mbox{c.c.}=0~. \lb{sdn}
\eea
One can readily show that, after making use of the relation \p{FVrel},
these conditions can be brought into the form quite similar to the $U(1)$
self-duality condition \p{newsd}
\bea
&& V^k_\ab\frac{\partial E}{\partial V_\ab^l}-
\bar{V}^k_{\da\db}\frac{\partial E}{\partial\bar{V}_{\da\db}^l}=0~.
\lb{EUinvar}
\eea
This constraint is none other than the condition of invariance of
$E(V, \bar V)$ with respect to the $U(n)$ transformations \p{Htran}.

Using the $U(n)$-invariant function $E$ one can also construct
the simple invariant
\bea
&& I_0=-V^k_\ab\frac{\partial E}{\partial V_\ab^k}-
\bar{V}^k_{\da\db}\frac{\partial E}{\partial\bar{V}_{\da\db}^k}~.
\eea
It is easy to see that the whole Lagrangian \p{UVF} is not invariant with
respect to the nonlinear part of the $U(n)$ transformations, being
invariant only under the off-shell $SO(n)$ ones (corresponding to
$\eta^{kl} = 0$ in \p{Htran}). This matches with the fact that these
$U(n)/SO(n)$ transformations define an on-shell symmetry of the
joint set of equations of motion and Bianchi identities. Just these
transformations are true analogs of the standard $U(1)$ ($SO(2)$) duality
rotations discussed in previous Sections.

The conclusion is that the $V, F$ representation in the case of $n$
Abelian gauge field strengths allows one to reduce the nonlinear $U(n)$
self-duality conditions to the $U(n)$ invariance condition \p{EUinvar}
for the interaction function $E(V, \bar V)$ and so to obtain a general
description of the $U(n)$ self-dual models of nonlinear electrodynamics
in terms of an $U(n)$ invariant function of $n$ complex auxiliary
variables $V_\ab^i, \bar V^i_{\da\db}$.

The passing to the standard $F, \bar F$ representation of the
corresponding Lagrangians requires solving the nonlinear algebraic
equations \p{FVrel} for $V_\ab^i, \bar V^i_{\da\db}$. In general, this
can be done only within power expansions. It would be interesting
to find an example of self-dual system with a few gauge field
strengths, where a solution to such equations can be found
in a closed form, like in the BI example of Sect. 4, and to
establish the precise connection with the Lagrangian \p{Unrep}.

Finally, let us notice that the condition of ``discrete'' self-duality in
the general case is as follows
\bea
&& E(V^k_\ab, \bar V^k_{\da\db}) = E(-iV^k_\ab, i \bar V^k_{\da\db})~.
\lb{discrn}
\eea
It is an obvious generalization of the $n=1$ condition \p{discr1}.

\section{Conclusion}

This talk is an extended version of Sect. 3 of our work \cite{IZ}
devoted to the construction of $N=3$ supersymmetric Born-Infeld theory
(see also \cite{Ivtalk}). We have introduced a new $V, F$ representation
for the Lagrangians of nonlinear electrodynamics and shown that it allows
for a simple description of systems exhibiting the properties of on-shell
$U(n)$ self-duality or/and off-shell discrete self-duality in terms of
real function of auxiliary bispinor complex fields. In the $V, F$
representation, the nonlinear self-duality conditions are reduced to the
simple $U(1)$ or $U(n)$ invariance conditions \p{newsd}, \p{EUinvar} for
this function. The general condition of the discrete self-duality, or
self-duality under Legendre transformation, is the invariance of this
function under some discrete reflections of its arguments,
Eqs. \p{discr1}, \p{discrn}.

It is an interesting and quite feasible task to extend this consideration
to the case of $N=1$ and $N=2$ supersymmetric extensions of nonlinear
electrodynamics \cite{KT,ABMZ} in order to obtain a general
characterization of the corresponding self-dual systems. Also, finding
out a similar $V, F$  representation for non-Abelian BI theory and
its superextensions could shed more light on the structure of these
theories.
\vspace{0.3cm}

\noindent{\bf Note added.} Our general solution for the $SO(2)$ self-dual
Lagrangian $L(F^2,\bar F^2)$ can be obtained from Eq.\p{base2} by
substituting the $U(1)$-invariant function $E_{sd}(\nu,\bnu)=
\tilde{E}(\nu\bnu)$ and solving the algebraic equations \p{FV} for 
$V_{\alpha\beta}$. A similar form of the general solution to 
the $SO(2)$ self-duality condition has been found in Refs.\cite{GZ2} 
by a different method. To the best of our knowledge, the general solution 
of the $U(n)$ self-duality equation via a $U(n)$ invariant function $E$, 
as well as the general parametrization 
of the Lagrangians with discrete self-dualities via the functions 
$E$ \p{discr1}, \p{discrn} were not earlier considered in the literature.
\vspace{0.5cm}

\noindent{\bf Acknowledgements}
\vspace{0.3cm}

\noindent We are grateful to S.M. Kuzenko for bringing Refs.\cite{GZ2}
to our attention and for discussions. This work was partially supported by 
INTAS grant No 00-254, RFBR-DFG grant No 02-02-04002, grant DFG 436 RUS 
113/669, RFBR-CNRS grant No 01-02-22005 and a grant of Bogoliubov-Infeld
Programme.

\end{document}